\documentclass[cits]{PoS}
\usepackage{graphicx}

\newcommand{\be}{\begin{eqnarray}}
\newcommand{\ee}{\end{eqnarray}}

\newcommand\pt{p_{\rm \scriptscriptstyle T}}

\newcommand{\AmS}{{\protect\the\textfont2
  A\kern-.1667em\lower.5ex\hbox{M}\kern-.125emS}}

\title{QCD Predictions for Charm and Bottom Production at RHIC}

\ShortTitle{QCD Predictions for Charm and Bottom Production at RHIC}

\author{\speaker{Matteo Cacciari}\thanks{In collaboration with 
Paolo Nason and Ramona Vogt.}\\
        LPTHE - Universit\'e P. et M. Curie, France\\
	        E-mail: \email{cacciari@lpthe.jussieu.fr}}
		
\abstract{
We present up-to-date QCD predictions for open charm and bottom production at
RHIC in nucleon-nucleon collisions at $\sqrt{S} = 200$~GeV. The electron
spectrum resulting from heavy flavor decays is also evaluated for
direct comparison to the PHENIX and STAR data. These predictions seek to
establish a rigorous benchmark, including the theoretical uncertainties,
against which nuclear collision data can be compared to obtain evidence for
nuclear effects.
}

\FullConference{International Europhysics Conference on High Energy Physics\\
		 July 21st - 27th 2005\\
		 Lisboa, Portugal}
		 
\begin{document}

The PHENIX~\cite{Adler:2005fy} and
STAR~\cite{Adams:2004fc,Tai:2004bf} Collaborations at the Relativistic Heavy Ion
Collider (RHIC) have recently presented data for production of heavy quarks in 
$pp$ and d+Au collisions at $\sqrt{S_{NN}} = 200$ GeV, both in the form of explicitly
reconstructed $D$ mesons, and as an electron spectrum from semi-leptonic 
decay of
the heavy hadrons. These data must be compared to QCD predictions, 
to establish the extent to which they can be successfully described, before moving
forward and trying to determine the presence of dense matter effects in
nucleus-nucleus collisions.

Recent improvements in heavy quark production theory and experimental
measurements at colliders, especially for bottom production, have shown
that the perturbative QCD framework works rather well,
see Refs.~\cite{Cacciari:2004ur}, provided a number of
precautions are taken in performing the phenomenological analysis and the
proper modern tools are employed. In particular, the advantages of comparing
the theoretical prediction and the data directly at the level of
the experimental observables have been clearly outlined.

The purpose of this work is to repeat these analysis for  RHIC, and to
provide therefore a benchmark prediction against which the data can be compared.
 
We calculate the transverse momentum ($\pt$) distributions of
charm and bottom quarks, the charm and bottom hadron distributions
resulting from fragmentation and,
finally, the electrons produced in semi-leptonic decays of the hadrons
\cite{Cacciari:2005rk}.  Theoretical
uncertainties, estimated as extensively as possible, constitute an
intrinsic component of the prediction.
Our final result is thus not a single curve, but rather an
uncertainty band which has a reasonably large probability of containing
the `true' theoretical prediction.
 
The theoretical prediction of the electron spectrum includes
three  main components: the $\pt$ and rapidity
distributions of the heavy quark $Q$ in $pp$ collisions at $\sqrt{S} =
200$~GeV, calculated in perturbative QCD; fragmentation of the
heavy quarks into heavy hadrons, $H_Q$, described by
phenomenological input extracted from $e^+e^-$ data; and the decay of
$H_Q$ into electrons according to spectra available from other
measurements. This
cross section is schematically written as
\begin{eqnarray}
\frac{E d^3\sigma(e)}{dp^3} &=& \frac{E_Q d^3\sigma(Q)}{dp^3_Q} \otimes 
D(Q\to H_Q) \otimes f(H_Q \to e) 
\nonumber
\end{eqnarray}
where the symbol $\otimes$ denotes a generic convolution. 
%The electron decay
%spectrum, $f(H_Q \to e)$, accounts for the branching ratios.
 
The distribution $E d^3\sigma(Q)/dp^3_Q$ is evaluated at Fixed-Order plus
Next-to-Leading-Log (FONLL) level, implemented in Ref.~\cite{Cacciari:1998it}.
In addition to including the full fixed-order NLO
result~\cite{Nason:1987xz}, the FONLL calculation also
resums~\cite{Cacciari:1993mq} large perturbative terms proportional to
$\alpha_s^n\log^k(\pt/m)$ to all orders with next-to-leading logarithmic (NLL)
accuracy (i.e. $k=n,\,n-1$) where $m$ is the heavy quark mass. The perturbative
parameters are $m$ and the value of the strong coupling,
$\alpha_s$. We take $m_c = 1.5$~GeV and $m_b = 4.75$~GeV as central values
and vary the masses in the range $1.3< m_c <1.7$ GeV for charm and $4.5<m_b<
5$ GeV for bottom to estimate the mass uncertainties. The five-flavor QCD scale
is the CTEQ6M value, $\Lambda^{(5)} = 0.226$ GeV. 
The perturbative calculation also depends on the
factorization ($\mu_F$) and renormalization ($\mu_R$) scales.  The
scale sensitivity is a measure of the
perturbative uncertainty. We take
$\mu_{R,F} = \mu_0 = \sqrt{\pt^2 + m^2}$ as the central value and vary
the two scales independently within a  `fiducial' region defined by  $\mu_{R,F}
= \xi_{R,F}\mu_0$ with $0.5 \le \xi_{R,F} \le 2$ and $0.5 \le  \xi_R/\xi_F \le
2$ so that $\{(\xi_R,\xi_F)\}$ =
\{(1,1),  (2,2), (0.5,0.5), (1,0.5), (2,1), (0.5,1), (1,2)\}.  The envelope
containing the resulting curves defines the uncertainty.  The
mass and scale uncertainties are then added in quadrature.
These inputs lead to a FONLL  total $c\bar c$ cross section in
$pp$ collisions of $\sigma_{c\bar c}^{\rm FONLL} = 256^{+400}_{-146}$~$\mu$b
at $\sqrt{S} = 200$~GeV, and to a total $b\bar b$ cross section of 
$1.87^{+0.99}_{-0.67}$~$\mu$b.

% The theoretical uncertainty is evaluated as described
% above. The corresponding NLO  prediction is $244^{+381}_{-134}$~$\mu$b.
% The predictions in Ref.~\cite{Vogt:2001nh}, using $m_c = 1.2$~GeV and 
% $\mu_R = \mu_F = 2\sqrt{\pt^2 + m^2}$ gives
% $\sigma_{c\bar c}^{\rm NLO} = 427$~$\mu$b, within the uncertainties.
% Since the FONLL and NLO calculations tend to coincide at small $\pt$, which
% dominates the total cross section, the two results are very
% similar.
% Thus the two calculations are equivalent at the total cross section level,
% within the large perturbative uncertainties.  The total cross
% section for bottom production is $\sigma_{b\bar b}^{\rm FONLL} =
% 1.87^{+0.99}_{-0.67}$~$\mu$b.
 
\begin{figure}[t]
\begin{center}
\includegraphics[height=5cm]{cD.ps}~~~~~~
\includegraphics[height=5cm]{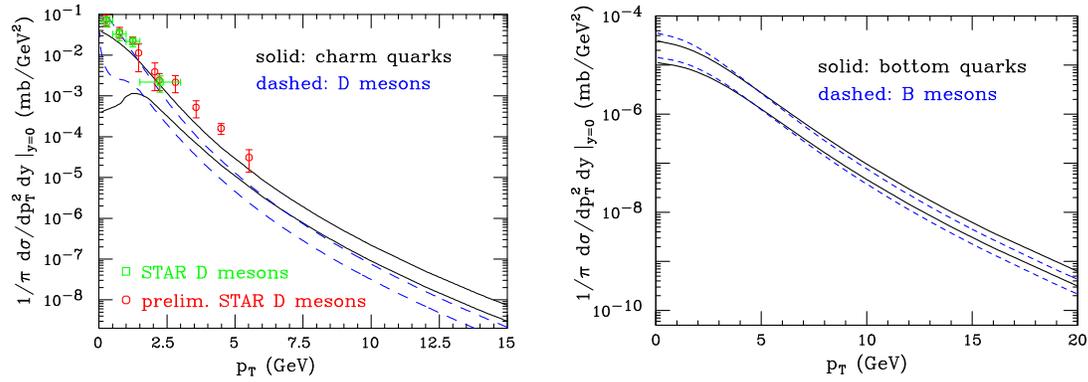}
\caption{\label{qQ} Left-hand side:  The theoretical uncertainty bands for 
$c$ quark and $D$ meson $p_T$ distributions in $pp$ collisions
at $\sqrt{S} = 200$~GeV, using BR($c \to D$) = 1. The 
final~\cite{Adams:2004fc} and preliminary~\protect\cite{Tai:2004bf} 
STAR d+Au data (scaled to $pp$ using $N_{\rm bin}$ =
7.5)
are also shown.  Right-hand side: The same for $b$ quarks and $B$
mesons.
}
\end{center}
\end{figure}
 
The fragmentation functions, $D(c\to D)$ and $D(b\to B)$, where $D$ and $B$
indicate a generic admixture of charm and bottom hadrons, are consistently
extracted from $e^+e^-$ data in the context of 
FONLL~\cite{Cacciari:2002pa,Cacciari:2005uk}.
Using the Peterson {\it et  al.} fragmentation
function, with standard parameter choices $\epsilon_c
\simeq
0.06 \pm 0.03$ and $\epsilon_b \simeq 0.006 \pm 0.003$, does not provide
a valid description of fragmentation in FONLL, since the hadronization is too
soft. In fact, Ref.~\cite{Cacciari:2002pa} showed 
that replacing the Peterson fragmentation description with an appropriate
one constitutes one of the main improvements 
which help reconcile the bottom transverse momentum distribution measured at 
the Tevatron with the theoretical prediction.

The measured spectra for primary $B\to e$ and $D \to e$ decays are modeled and
assumed to be equal for all bottom and charm hadrons, respectively.
 The contribution of
 electrons from secondary $B$ decays, $B\to D\to e$, was also included but is
 mostly negligible.
% was obtained
% by convoluting the $D\to e$ spectrum with a parton-model prediction of 
% $b\to c$ decay.  The resulting electron spectrum is very soft, giving a 
% negligible contribution to the total.
The decay spectra are normalized using the branching ratios
for bottom and charm hadron mixtures:
BR$(B\to e) = 10.86 \pm 0.35$\%, BR$(D\to e) = 10.3 \pm 1.2$\%,
and BR$(B\to D\to e) = 9.6 \pm 0.6$\%.
 
The left-hand side of Fig.~\ref{qQ}
shows the theoretical uncertainty bands for $c$ quarks and $D$ mesons, 
obtained by summing the mass and scale uncertainties
in quadrature.  The band is broader at low
$\pt$ due to the large value of $\alpha_s$ and the behavior of the CTEQ6M
parton densities at low
scales as well as the increased sensitivity of the cross section to the
charm quark mass.  
% The rather hard
% fragmentation function causes the $D$ meson and $c$ quark bands to
% separate only at $\pt > 9$~GeV.
% The right-hand side of Fig.~\ref{qQ} shows the same results for $b$ quarks and
% $B$ mesons.  The even harder $b\to B$ fragmentation
% function causes the two bands to partially overlap until
% $\pt \simeq 20$~GeV.
 
\begin{figure}[t]
\begin{center}
\includegraphics[height=5cm]{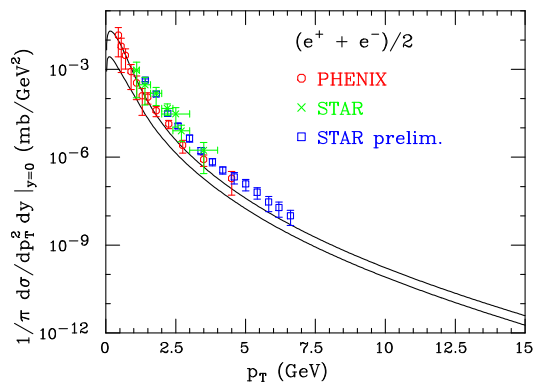}
\caption{\label{electrons} 
The final prediction for the
theoretical uncertainty band of the electron spectrum from charm and
bottom in $pp$ collisions. 
Data from PHENIX~\protect\cite{Adler:2005fy} 
and STAR (final~\protect\cite{Adams:2004fc} and preliminary~\protect\cite{Tai:2004bf})
are also shown.}
\end{center}
\end{figure}

Figure~\ref{electrons} 
 compares the RHIC data to the total uncertainty band for 
$D\to e$, $B \to e$ and $B\to D \to e$ decays to electrons. The 
upper and lower limits of the
band are obtained by summing the upper and lower limits for each component.
It is worth noting that, while for the central parameter
sets, the $B \to e$ decays begin to dominate the $D \to e$ decays
at $\pt \simeq 4$~GeV, a comparison of the individual bands (not shown) 
shows that the crossover may
occur over a rather broad range of electron $\pt$.  The relative $c$ and $b$
decay contributions may play an important part in understanding the electron
$R_{AA}$ in nucleus-nucleus collisions 
%\cite{STARaa,PHENIXaa} 
which seems to
suggest strong energy loss effects on heavy flavors \cite{Magda,Nestor}.

In conclusion, we have performed a phenomenological analysis of heavy quark
production  in $\sqrt{S} = 200$~GeV $pp$ collisions at RHIC.  The
results are presented in the form of a theoretical uncertainty band for the
transverse momentum distribution of  either bare charm (bottom), $D$ ($B$)
mesons, or electrons originating from the decay of charm and bottom hadrons.
This band stems, at the perturbative level, from a next-to-leading order result
improved by next-to-leading log resummation at large transverse momenta.
These result should
not be multiplied by any $K$ factor before comparison with data. Rather,
agreement within the uncertainties of the measurements will support the
applicability of standard QCD calculations to heavy quark production at RHIC.
Alternatively, a significant disagreement will suggest the need to complement
this evaluation with further ingredients.


\begin{thebibliography}{99}
 
%\cite{Adler:2005fy}
\bibitem{Adler:2005fy}
  S.~S.~Adler {\it et al.}  [PHENIX Collaboration],
  %``Single electrons from heavy flavor decays in p + p collisions at s**(1/2) =
  %200-GeV,''
  arXiv:hep-ex/0508034.
  %%CITATION = HEP-EX 0508034;%%

 
%%\cite{Kelly:2004qw}
%\bibitem{Kelly:2004qw}
%S.~S. Adler {\it et al.} [PHENIX Collaboration],
%%``The PHENIX measurement of heavy flavor via single electrons in p p, d - Au,
%%and Au - Au collisions at s(NN)**(1/2) = 200-GeV,''
%arXiv:hep-ex/0508034.
%%%CITATION = NUCL-EX 0403057;%%
 

%\cite{Adams:2004fc}
\bibitem{Adams:2004fc}
  J.~Adams {\it et al.}  [STAR Collaboration],
  %``Open charm yields in d + Au collisions at s(NN)**(1/2) = 200-GeV,''
  Phys.\ Rev.\ Lett.\  {\bf 94} (2005) 062301
  [arXiv:nucl-ex/0407006].
  %%CITATION = NUCL-EX 0407006;%% 
  
  
%\cite{Tai:2004bf}
\bibitem{Tai:2004bf}
A.~Tai  [STAR Collaboration],
%``Measurement of open charm production in d + Au collisions at s(NN)**(1/2) =
%200-GeV,''
J.\ Phys.\ G {\bf 30} (2004) S809
[arXiv:nucl-ex/0404029].
%%CITATION = NUCL-EX 0404029;%%

%\cite{Cacciari:2004ur}
\bibitem{Cacciari:2004ur}
%\cite{Mangano:2004xr}
%\bibitem{Mangano:2004xr}
M.~L.~Mangano,
arXiv:hep-ph/0411020;
%%CITATION = HEP-PH 0411020;%%
M.~Cacciari,
arXiv:hep-ph/0407187.
%%CITATION = HEP-PH 0407187;%%
% 

%\cite{Cacciari:2005rk}
\bibitem{Cacciari:2005rk}
  M.~Cacciari, P.~Nason and R.~Vogt,
  %``QCD predictions for charm and bottom production at RHIC,''
  Phys.\ Rev.\ Lett.\  {\bf 95} (2005) 122001
  [arXiv:hep-ph/0502203].
  %%CITATION = HEP-PH 0502203;%%

 
%\cite{Cacciari:1998it}
\bibitem{Cacciari:1998it}
M.~Cacciari, M.~Greco and P.~Nason,
%``The p(T) spectrum in heavy-flavour hadroproduction,''
JHEP {\bf 9805} (1998) 007
[arXiv:hep-ph/9803400];
%%CITATION = HEP-PH 9803400;%%
%\cite{Cacciari:2001td}
%\bibitem{Cacciari:2001td}
M.~Cacciari, S.~Frixione and P.~Nason,
%``The p(T) spectrum in heavy-flavor photoproduction,''
JHEP {\bf 0103} (2001) 006
[arXiv:hep-ph/0102134].
%%CITATION = HEP-PH 0102134;%%
 
%\cite{Nason:1987xz}
\bibitem{Nason:1987xz}
P.~Nason, S.~Dawson and R.~K.~Ellis,
%``The Total Cross-Section For The Production Of Heavy Quarks In Hadronic
%Collisions,''
Nucl.\ Phys.\ B {\bf 303} (1988) 607;
%%CITATION = NUPHA,B303,607;%%
%\cite{Nason:1989zy}
%\bibitem{Nason:1989zy}
P.~Nason, S.~Dawson and R.~K.~Ellis,
%``The One Particle Inclusive Differential Cross-Section For Heavy Quark
%Production In Hadronic Collisions,''
Nucl.\ Phys.\ B {\bf 327} (1989) 49
[Erratum \ B {\bf 335} (1990) 260];
%%CITATION = NUPHA,B327,49;%%
%\cite{Beenakker:1990ma}
%\bibitem{Beenakker:1990ma}
W.~Beenakker, W.~L.~van Neerven, R.~Meng, G.~A.~Schuler and J.~Smith,
%``QCD Corrections To Heavy Quark Production In Hadron Hadron Collisions,''
Nucl.\ Phys.\ B {\bf 351} (1991) 507.
%%CITATION = NUPHA,B351,507;%%
 
%\cite{Cacciari:1993mq}
\bibitem{Cacciari:1993mq}
M.~Cacciari and M.~Greco,
%``Large p(T) hadroproduction of heavy quarks,''
Nucl.\ Phys.\ B {\bf 421} (1994) 530
[arXiv:hep-ph/9311260].
%%CITATION = HEP-PH 9311260;%%

%\cite{Vogt:2001nh}
% \bibitem{Vogt:2001nh}
% R.~Vogt,
% %``The A dependence of open charm and bottom production,''
% Int.\ J.\ Mod.\ Phys.\ E {\bf 12} (2003) 211
% [arXiv:hep-ph/0111271].
% %%CITATION = HEP-PH 0111271;%%
 
%\cite{Cacciari:2002pa}
\bibitem{Cacciari:2002pa}
M.~Cacciari and P.~Nason,
%``Is there a significant excess in bottom hadroproduction at the  Tevatron?,''
Phys.\ Rev.\ Lett.\  {\bf 89} (2002) 122003
[arXiv:hep-ph/0204025].
%%CITATION = HEP-PH 0204025;%%
 
%\cite{Cacciari:2005uk}
\bibitem{Cacciari:2005uk}
  M.~Cacciari, P.~Nason and C.~Oleari,
  %``A study of heavy flavoured meson fragmentation functions in e+ e-
  %annihilation,''
  arXiv:hep-ph/0510032.
  %%CITATION = HEP-PH 0510032;%% 
 
% %\cite{Peterson:1982ak}
% \bibitem{Peterson:1982ak}
% C.~Peterson, D.~Schlatter, I.~Schmitt and P.~M.~Zerwas,
% %``Scaling Violations In Inclusive E+ E- Annihilation Spectra,''
% Phys.\ Rev.\ D {\bf 27} (1983) 105.
% %%CITATION = PHRVA,D27,105;%%
 
% %\cite{Eidelman:2004wy}
% \bibitem{Eidelman:2004wy}
% S.~Eidelman {\it et al.}  [Particle Data Group Collaboration],
% %``Review of particle physics,''
% Phys.\ Lett.\ B {\bf 592} (2004) 1.
% %%CITATION = PHLTA,B592,1;%%

%\bibitem{STARaa} J. Bielcik [STAR Collaboration], these proceedings.

%\bibitem{PHENIXaa} S. A. Butsyk [PHENIX Collaboration], these proceedings.

%\bibitem{Magda} M. Djordjevic, these proceedings.
%\cite{Djordjevic:2005db}
\bibitem{Magda}
  M.~Djordjevic, M.~Gyulassy, R.~Vogt and S.~Wicks,
  %``Influence of bottom quark jet quenching on single electron tomography of Au
  %+ Au,''
  arXiv:nucl-th/0507019.
  %%CITATION = NUCL-TH 0507019;%%
  
%\cite{Armesto:2005mz}
\bibitem{Nestor}
  N.~Armesto, M.~Cacciari, A.~Dainese, C.~A.~Salgado and U.~A.~Wiedemann,
  %``How sensitive are high-p(T) electron spectra at RHIC to heavy quark energy
  %loss?,''
  arXiv:hep-ph/0511257.
  %%CITATION = HEP-PH 0511257;%%
%\bibitem{Nestor} N. Armesto, these proceedings. 
\end{thebibliography}
\end{document}